\documentclass[twocolumn]{aastex62}
\usepackage{color}
\usepackage{appendix}

\newcommand{\kms}{\ifmmode {\rm km\ s}^{-1} \else km s$^{-1}$\ \fi}
\newcommand{\ergs}{\ifmmode {\rm erg\ s}^{-1} \else erg s$^{-1}$\ \fi}

\newcommand{\feii}{Fe {\sc ii}\ }
\newcommand{\mgii}{Mg {\sc ii}}
\newcommand{\civ}{C {\sc iv}}

\newcommand{\lb}{\ifmmode L_{\rm Bol} \else $L_{\rm Bol}$\ \fi}
\newcommand{\ledd}{\ifmmode L_{\rm Edd} \else $L_{\rm Edd}$\ \fi}
\newcommand{\lx}{\ifmmode L_{\rm 2-10keV} \else  $L_{\rm 2-10keV}$\ \fi}
\newcommand{\hb}{\ifmmode H\beta \else H$\beta$\ \fi}
\newcommand{\ha}{\ifmmode H\alpha \else H$\alpha$\ \fi}
\newcommand{\heii}{He {\sc ii}\ }
\newcommand{\oiii}{[O {\sc iii}]\ }
\newcommand{\oii}{[O {\sc ii}]\ }

\newcommand{\mbh}{\ifmmode M_{\rm BH}  \else $M_{\rm BH}$\ \fi}
\newcommand{\lv}{\ifmmode \lambda L_{\lambda}(5100\AA) \else $\lambda L_{\lambda}(5100\AA)$\ \fi}
\newcommand{\msun}{M_{\odot}}
\newcommand{\mdot}{\ifmmode \dot{m} \else \dot{m} \fi }
\newcommand{\llog}{\ifmmode {\rm log} \else {\rm log} \fi }
\newcommand\gtsima{$\; \buildrel > \over \sim \;$}
\newcommand\simgt{\lower.5ex\hbox{\gtsima}}

\def\cm2{{cm$^{-2}$}}

\newcommand{\leddR}{\ifmmode L_{\rm Bol}/L_{\rm Edd} \else $L_{\rm Bol}/L_{\rm Edd}$\ \fi}

\newcommand{\aox}{$\alpha_{\rm ox}$}

\graphicspath{{./}{figures/}}

\received{***}
\revised{***}
\accepted{***}
\submitjournal{AJ}

\shorttitle{The BLUESHIFT OF \civ}
\shortauthors{Ge et al.}

\begin{document}

\title{THE BLUESHIFT OF \civ\ BROAD EMISSION LINE IN QSOs}
\shortauthors{Ge et al.}

\correspondingauthor{Wei-Hao Bian}
\email{whbian@njnu.edu.cn}

\author[0000-0003-4490-9307]{Xue Ge}
\affiliation{School of Physics and Technology, Nanjing Normal University, Nanjing 210023, China}

\author[0000-0002-5413-3680]{Bi-Xuan Zhao}
\affil{School of Physics and Technology, Nanjing Normal University, Nanjing 210023, China}

\author{Wei-Hao Bian}
\affil{School of Physics and Technology, Nanjing Normal University, Nanjing 210023, China}

\author{Green Richard Frederick}
\affiliation{Steward Observatory, University of Arizona, Tucson, AZ 85721, USA}




\begin{abstract}

For the sample from Ge et al. of 87 low-$z$ Palomar--Green (PG) quasi-stellar objects (QSOs) and 130 high-$z$ QSOs ($0<z<5$) with $\hb$-based single-epoch supermassive black hole (SMBH) masses, we performed a uniform decomposition of the \civ\ $\lambda$1549 broad-line profile. Based on the rest frame defined by the \oiii $\lambda$5007 narrow emission line, a medium-strong positive correlation is found between the \civ\ blueshift and the luminosity at 5100\AA\  or the Eddington ratio \leddR. A medium-strong negative relationship is found between the \civ\ blueshift and \civ\ equivalent width. These results support the postulation where the radiation pressure may be the driver of \civ\ blueshift. There is a medium strong correlation between the mass ratio of \civ-based to
$\hb$-based \mbh and the \civ\ blueshift, which indicates that the bias for \civ-based \mbh is affected by the \civ\ profile. 

\end{abstract}



\section{Introduction} \label{sec:intro}

Broad emission lines are the most prominent spectral features of Type I active galactic nuclei (AGNs) and quasi-stellar objects (QSOs).  It is believed that broad emission lines are produced by photoionization. The ionizing photons are from the accretion disc surrounding the central supermassive black hole (SMBH) in AGNs/QSOs. These photons irradiate the high-velocity clouds in broad line regions (BLRs) and the subsequent recombination produces the broad emission lines. The broad emission lines have been studied for many years with regard to their geometry and kinematics. The \mbox{observed} properties of broad emission lines including the width, strength, and profile can give us insight into the physical processes in the central regions of AGNs/QSOs \citep[e.g.,][]{Wang2017}.

The blueshift of the \civ\ emission line relative to \mbox{low-ionization} lines is unambiguously detected in many samples \citep[e.g.,][]{Gaskell1982, Richards2002, Vanden Berk2001, Wang2011}. It indicates that an outflowing wind may be a common configuration for the BLR clouds \citep[]{Gaskell1982, Marziani1996, Leighly2004}. \cite{Richards2011} proposed a two-component model for BLRs, namely a wind and a disk component. In this model, a stronger EUV ionizing continuum will reduce the wind component relative to the disk lines by ionizing the atoms that would otherwise produce the line-driven wind; for weaker EUV/soft \mbox{X-ray}, the wind component can develop, and the disk component is \mbox{further} reduced because the wind absorbs some of the ionizing continuum. The \civ\ blueshift quantifies the relative strength of these two components. The \civ\ equivalent width (EW) quantifies the total BLR gas and ionizing UV photons reaching the disk.

Actually, the accuracy of the \civ\ blueshift determination depends on the measurement of systemic redshift. There are several ways used to obtain AGNs/QSOs redshifts. (1) Host galaxy absorption lines. However, these host absorption lines are usually contaminated by the light from the AGN/QSOs in spatially unresolved \mbox{spectra}. It makes the absorption lines too weak to be recognized. (2) The narrow \oiii or Balmer lines. These lines have been measured to have very low blueshifts or redshifts \citep[e.g.][]{Marziani1996, Sulentic2000, Richards2002, Hewett2010}. In more distant QSOs ($z > 0.8$), \oiii and \hb have shifted into the near-infrared (NIR), for which infrared spectrographs need be used to measure the common rest frame optical lines of these QSOs. (3) UV emission lines. \mgii\ $\lambda$2798 is accepted to be the most dependable UV emission line for the measurement of  AGNs/QSOs redshifts \citep[e.g.,][]{Tytler1992}. However, when $z > 2.2$, the wavelength of \mgii\ moves beyond the optical window. \civ\ $\lambda$1549 will be a reliable alternative for the measurement of redshift when \mgii\ is not available. However, the \civ\ outflow makes difficulties for the measurement of the redshifts.

The \hb line is widely adopted to estimate the \mbh of AGNs/QSOs \citep[e.g.,][]{Kaspi2000, Collin2006, VP06, Mclure2002, Onken2008, Ge2016} because of its calibration through reverberation mapping at low redshift. For the high-ionization lines, such as \civ, \cite{Baskin2004} found that, for QSOs with \hb FWHM $<$ 4000 \kms, the \mbox{\civ-based} \mbh is higher by a factor of 3-4 than the \mbox{$\hb$-based} \mbh. In contrast, for QSOs with \hb FWHM $>$ 4000 \kms, the \mbox{\civ-based} \mbh is lower by the same factor. The \mbox{\civ-based} \mbh is biased with respect to the \mbox{$\hb$-based} \mbh \citep[e.g.,][]{Bian2012,Shen2012}. The uncertainties in the determination of \mbh may arise from the non-virial component shown in the line profile, such as from a radiation driven disk wind \citep[e.g.,][]{Murray1995, Proga2000, Richards2011, Coatman2017}.

In this paper, for a compiled sample of low-$z$ Palomar-Green (PG) QSOs and high-$z$ QSOs, spectral decomposition is used to investigate the blueshift of the \civ\ broad emission line relative to \oiii $\lambda$5007. The sample and analysis are described in \S 2. Our result and discussion are given in \S 3. Finally, our conclusions are presented in \S 4. All of the cosmological calculations in this paper assume $H_{0}=70 \rm {~km ~s^ {-1}~Mpc^{-1}}$, $\Omega_{M}=0.3$, and $\Omega_{\Lambda} = 0.7$.

\section{Sample and Analysis}
\subsection{Sample selection}

A low-$z$ sample is adopted from the opticaly selected sample of PG QSOs \citep{BG92}. It contains 87 PG QSOs with $0<z<0.5$ from the Bright Quasar Survey \citep[]{Schmidt1983}. It is the most thoroughly explored sample of AGN/QSO, with a lot of high-quality broadband data, including X-ray, optical, infrared and radio \citep[e.g.,][]{BG92, Brandt2000, BL2004, Baskin2004, Shi2014, Bian2016}. The optical spectra of the 87 objects are from the Gold Spectrograph on the KPNO 2.1 m telescope. The resolution of the spectra is 6.5 \AA\ corresponding to $\sim$ 400 \kms. The redshift of each QSO was measured from the \oiii $\lambda$ 5007 narrow emission line except for 8 QSOs with weak or absent \oiii. Their redshifts were measured through the \hb line \citep[]{BG92}.

The UV spectra covering \civ\ 1549\AA\ of 85 PG QSOs are available in the MAST archive, 47 from Hubble Space Telescope (HST) and 38 from the International Ultraviolet Explorer (IUE). Three sources, PG 0934+013, PG 1004+130, and PG 1448+273, do not have enough S/N to fit their \civ\ profiles. We also exclude fifteen sources that are heavily absorbed in the \civ\ region \citep[]{Laor2002,Baskin2004} because our fitting method used here is not suitable for profiles with a broad absorption line (BAL). Finally, there are 67 PG QSOs which are available for spectral decomposition. The UV spectra for 35 PG QSOs are available from HST, which were observed with the Faint Object Spectrograph (FOS) covering the wavelength range $1150-8500$\AA. The spectral resolutions ($R=\lambda/\Delta \lambda$) for FOS observations are $R \approx 1300$ and $R \approx 250$ at high and low spectral resolution, respectively. The UV spectra for the 32 PG QSOs are available from IUE with spectral resolution of $R \approx 200$ \citep[]{Anand2009}. The wavelength range of the IUE UV spectra are $1151-2000$ \AA\ and $1850-3400$ \AA\ at short wavelength Prime (SWP) and Long wavelength Prime (LWP) respectively. 
For some PG QSOs with IUE spectra, we did not add other observations, although a small number of sources were observed again using other spectrographs. We think that the choice spectrographs of  will not impact our results. Table \ref{Table 1} lists the information for these 67 PG QSOs. 

The SDSS I/II data (Data Relaease from 1 to 7, DR1-DR7) are available from the dedicated wide-field 2.5 m telescope \citep[]{Gunn2006} at Apache Point Observatory near Sacramento Peak in Southern New Mexico. The spectral resolution is up to 69 \kms. The wavelength range is from 3800 \AA\ to 9200 \AA. SDSS III (Data Relaease from 8 to 12, DR8-DR12) extends the range of wavelength from 3600 \AA\ to 10400 \AA.
\cite{Ge2016} collected 181 high-$z$ QSOs with $\hb$-based \mbh from different literatures. They crossed match the sample with \cite{Shen2011} and obtained only 125 sources with \civ\ fitting data \footnote{SDSS081331+254503 and HS0810+2554 in their Table 2 are actually the same.}. In this paper, we use 130/181 sources where 125 objects are from \cite{Ge2016} and their UV spectra are obtained from SDSS DR7 \citep[]{Abazajian2009, York2000}. For 5 of the 130 high-$z$ QSOs (i.e., TON618, UM667, SDSSJ105123+35453, SDSSJ153650+50081 and SDSSJ165354+40540), their UV spectra are from SDSS DR12 \citep[]{Paris2016} because the UV spectra with \civ\ are not available in SDSS DR7.

\begin{figure*}
\centering
\includegraphics[width=0.5\textwidth]{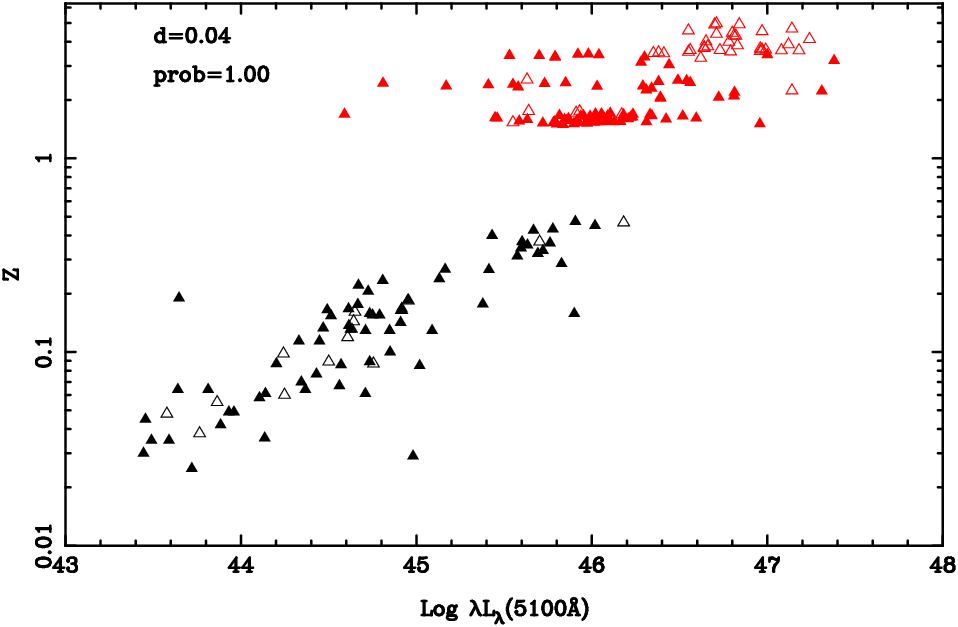}
\caption{The distribution of low-$z$ QSOs (black) and high-$z$ QSOs (red) in $z$-\lv space.The hollow triangles are excluded from our sample. The labels in the panel are the two-dimension K-S test. }
\label{fig1}
\end{figure*}

To get reliable measurements of the blueshift of the \civ\ line, we constructed a subsample from these \mbox{low-$z$} and high-$z$ QSOs, i.e., 67 PG QSOs and 112 \mbox{high-$z$} QSOs. For 
high-$z$ QSOs, we eliminated 9 BAL QSOs identified by \cite{Shen2011} and visualizing and 7 weak-line sources from \cite{Shemmer2015}. In addition, we also excluded Q0142-100 and PG 1115+080 that are from \cite{Assef2011}, because of the lack of a redshift estimate from \oiii or \mbox{H$\beta$/\ha}. The high-$z$ QSOs in subsample comprises 112 QSOs, where 60 are from \cite{Shen2012}, 3 are from \cite{Assef2011}, 12 are from \cite{Shemmer2004}, 15 are from \cite{Netzer07} and 22 are from \cite{Jun2015}. We perform the two-dimensional K-S test on the entire sample and subsample and find no significant difference between the two samples (Figure \ref{fig1}). The large p value ($>0.001$) shows that our selection of subsample does not significantly affect the results of this work.  We call the subsample as ``sample'' which will be analysed in following sections.
Table \ref{Table 2} lists the information for these high-$z$ QSOs.

The high-$z$ sources were chosen with high S/N in SDSS spectra, and have H$\beta$/\oiii fall in a good place in the near infrared atmospheric window. The high S/N, however, may lead the QSOs with very broad lines and low EW in \civ\ to be under-represented in this work. One property of our sample is that PG and high-$z$ QSOs are bright in UV band. This property reduces the objects that are reddened by circum-nuclear dust might differentially suppresses some of the \civ\ emitting material. The dust-obscured objects may lead a different correlation, but would not be as valid for learning about the physical origin of the total \civ\ profile.
It is worth emphasizing that the \mbh of the sample is H$\beta$-based thanks to the infrared observation. This unity enables us to explore the relationship between the \civ\ blueshift and \leddR from low-$z$ to high-$z$ QSOs. In addition, There are 14 and 11 radio loudness QSOs in PG QSOs and high-$z$ QSOs respectively. We do not exclude these objects and will discuss their \civ\ blueshift in Section 3.

\subsection{UV spectral decomposition}

\begin{figure*}
\centering
\includegraphics[width=0.3\textwidth]{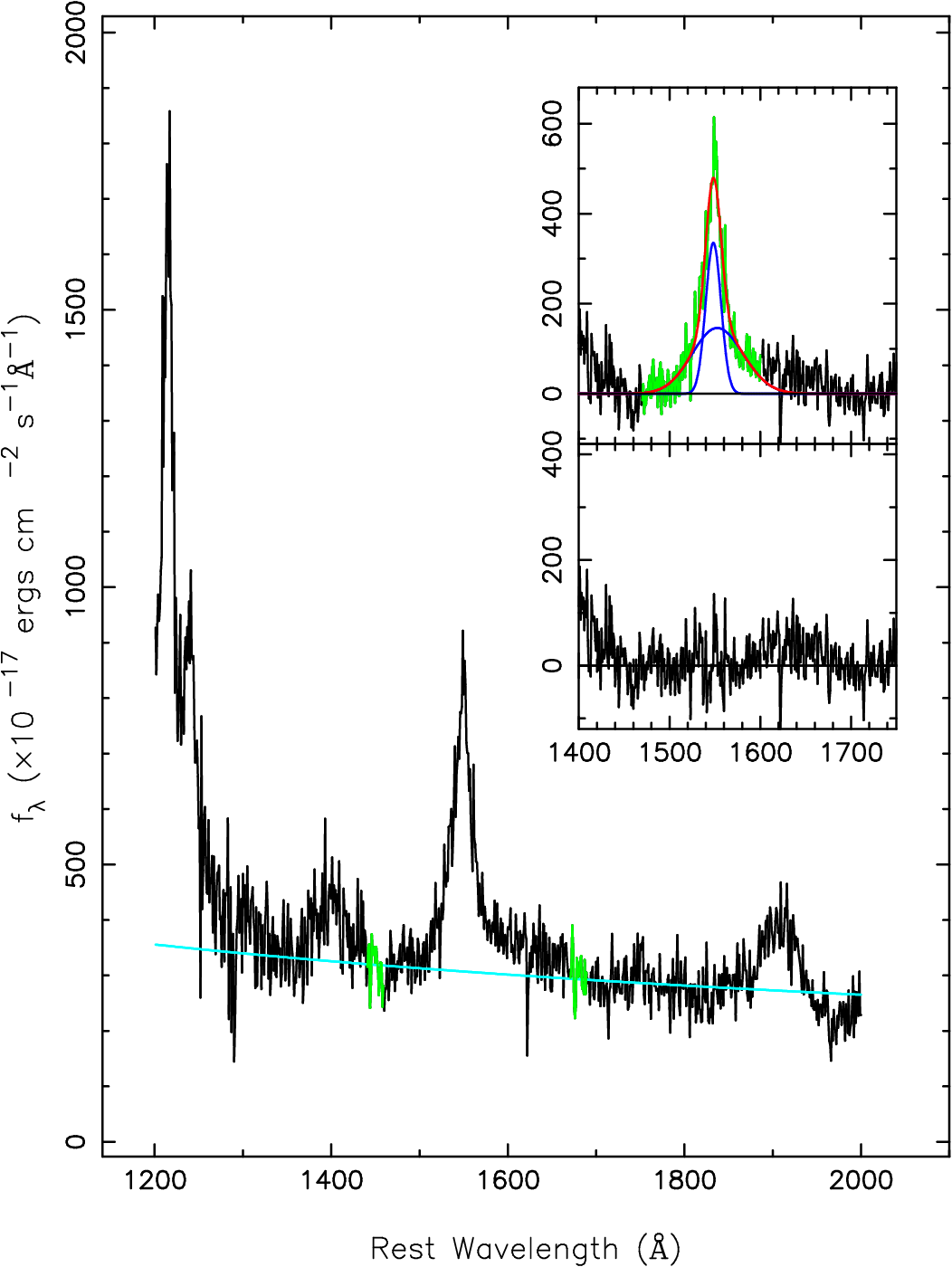}
\includegraphics[width=0.3\textwidth]{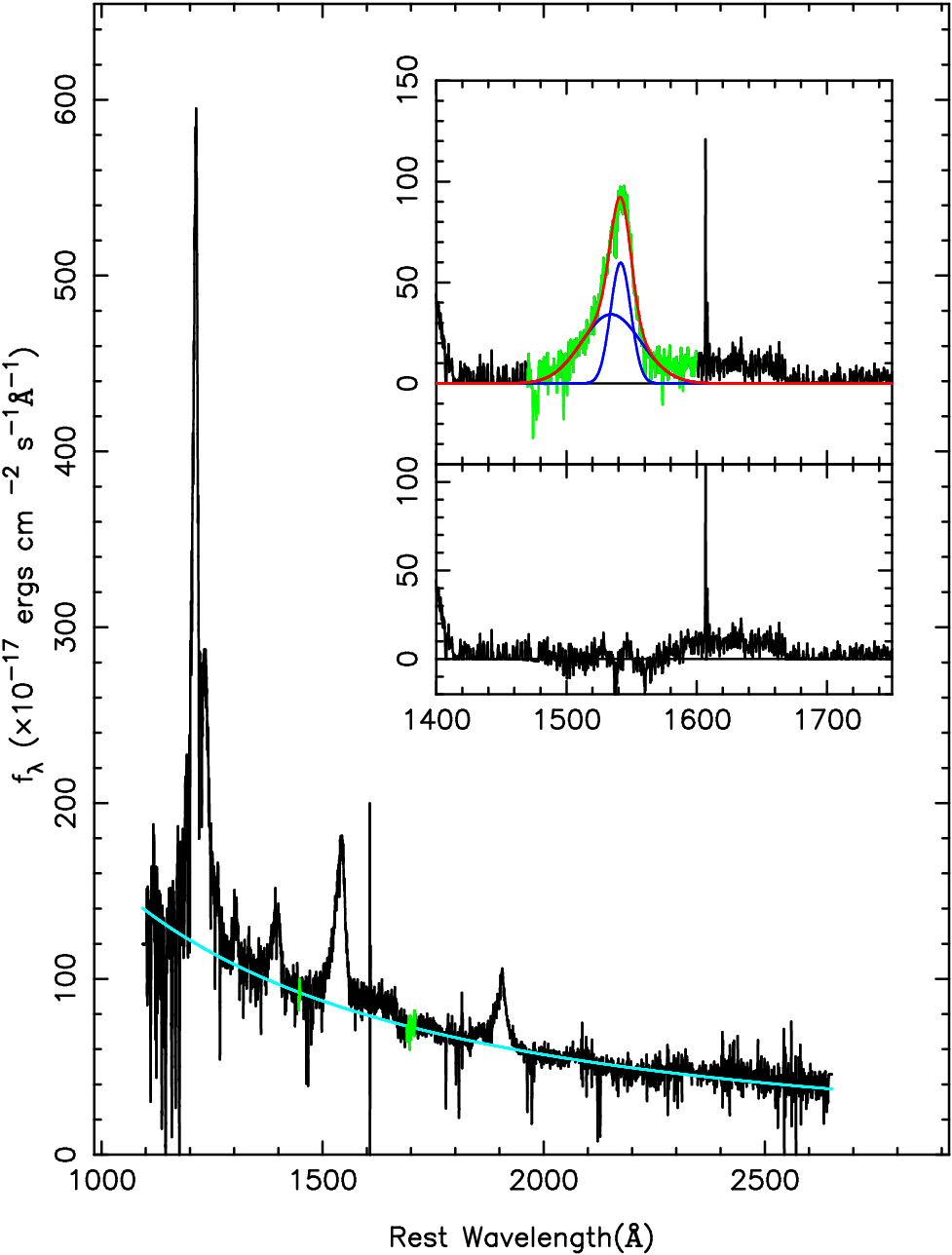}
\caption{Examples of the UV continuum and emission line fittings for a low-$z$ (left) and a high-$z$ (right) QSO. The original spectra and the fitting windows for the continuum are marked in black and green. The cyan curve is the power law which represents the continuum. The inset shows the fitting of the \civ\ emission line. The fitting window is also shown in green. The narrow and broad components are marked in blue, and their sum is marked in red. The residual spectrum is shown at the bottom.}
\label{fig2}
\end{figure*}

We need to point out that an average spectrum, weighted by the S/N ratio, is adopted when more than one spectrum is available.
We performed least-$\chi^{2}$ fits to the UV emission line spectra. The $\chi^{2}$ is determined by the error in the flux. The mean reduced $\chi^{2}$ are 1.43 and 1.25 for high-$z$ and low-$z$ QSOs respectively. We list the steps of spectral decomposition as follows.
First, the spectra are corrected for Galactic extinction using their $A_V$ values from the NASA/IPAC Extragalactic Database assuming an extinction curve with $R_{V} = 3.1$ \citep[]{Cardelli1989}. The extinction-corrected observed spectra are then transformed to the rest frame; no extinction correction is applied for the host galaxy. For PG QSOs, we use the redshifts from \cite{BG92}. For high-$z$ QSOs, we adopt the redshifts from \cite{Hewett2010}. For UM667 and TON618, their redshifts are adopted from the NASA/IPAC Extragalactic Database because their redshifts are not available in \cite{Hewett2010}. Next, we fit the continuum using a power law function in two fitting windows near $\sim 1450$ \AA\ and $\sim 1710$ \AA. Finally, the \civ\ emission line in the continuum-subtracted spectrum is fitted with two Gaussians, one with intermediate velocity width and another with broader width. The FWHM of the two components from BLRs is restricted to be more than 1000   and less than 20000 \kms. Simultaneously, we constrain the centers of their profiles to have a shift range of $\sim$15 \AA. The fitting window for the \civ\ line covers the range of 1470-1600\AA\ considering the contamination of \heii $\lambda$ 1640. It is found that a two-Gaussian fit is adequate for the \civ\ profile in our low-$z$ and high-$z$ QSOs. Figure \ref{fig2} shows the examples of the fitting of the continuum and the \civ\ line. We list the profile parameters of low- and high-z QSOs in Tables 3 and 4.

\citet[]{Wills1993} and \citet[]{Brotherton1994} presented statistical investigation of broad emission-line profiles (such as \civ, \mgii) in QSOs. They proposed a two-component model where a narrow core component (FWHM $\sim$ 2000 \kms) and a broad base component (FWHM $\sim$ 7000 \kms) with blueshift $\sim$ 1000 \kms relative to the core component to interpret their results. They suggested that the core-to-base ratio determines the profile of emission line.

\subsection{\rm \civ\ BLUESHIFT, \mbh, and \leddR}
For the two-Gaussian fitting of the \civ\ line, the EW(\civ) of the entire profile is calculated by integrating both components and the results are consistent with \citet[]{Wu2009}. The error for EW(\civ) is derived from the errors of the integrated \civ\ flux and the continuum flux in our fits.

From the reconstructed \civ\ profile of our two-Gaussian fitting, the \civ\ peak wavelength is calculated, as well as its error. With respect to the rest frame defined by the \oiii emission line or the Balmer emission line (i.e., $z_{\rm sys}$), the \civ\ blueshift and the error can be calculated following Equations \ref{e1} and \ref{e2}, where $c$ is the speed of light. In Equation \ref{e1}, $\lambda$ is the \civ\ peak wavelength in the rest frame defined by $z_{\rm fit}$ used in our fitting, and $z_{\rm sys}$ is the systemic redshift that is defined by the \oiii or Balmer emission line. The error of the \civ\ blueshift is determined by the error of the peak wavelength and errors of these two redshifts, which is related to the spectral resolution during observations.

\begin{equation}
v=(\frac{\lambda(1+z_{fit})}{1+z_{sys}}-1549)\frac{c}{1549}
\label{e1}
\end{equation}

\begin{eqnarray}
(\delta v)^{2}=
((\frac{\delta\lambda(1+z_{fit})}{1+z_{sys}})^2+(\frac{\lambda\delta z_{fit}}{1+z_{sys}})^2+\nonumber\\
(\frac{\lambda\delta z_{sys}(1+z_{fit})}{(1+z_{sys})^2})^2)
\times (\frac{c}{1549})^{2} 
\label{e2}
\end{eqnarray}

We calculate the error of the \civ\ blueshift according to the error transfer formula (Equation 2). For PG QSOs, the resolution of the optical spectra is 6.5 \AA\, corresponding to $\sim$ 400 \kms. The resolution of the UV spectra is $\sim$ 460 \kms and $\sim$ 1200 \kms for HST and IUE data, respectively. For a high-$z$ QSO ($z=2$) with infrared spectral resolution R $\sim$ 100, the uncertainty of its blueshift is $\sim$ 1000 \kms. If R $\sim$ 3000, then the uncertainty of its blueshift is $\sim$ 30 \kms. For SDSS UV spectra, $R\sim 2000$, and the uncertainty of blueshifts is $\sim$ 50 \kms. Typically, the centroid of a well-shaped emission line can be determined to at least 0.1 or better of the spectral resolution. The larger uncertainty in the \civ\ blueshift comes from noise in the broad emission line profiles \citep[]{BG92}. We can therefore ignore the effect of spectral resolution on the error of the blueshift. For that reason, we do not consider the uncertainties from the spectral resolution and the two redshifts, $z_{fit}$ and $z_{sys}$ for either low- or high-$z$ QSOs. The error of the \civ\ blueshift is calculated only from the error of the fit of the \civ\ peak wavelength.

For our sample of low-$z$ and high-$z$ QSOs,  \cite{Ge2016} has computed the $\hb$-based single-epoch SMBH mass, \mbh and host-corrected \leddR. They estimated the bolometric luminosity based on a constant bolometric correction (9.26) and host-corrected $L (5100\AA)$ (see \cite{Ge2016}, for details) . We use these parameters to investigate the relation with the \mbox{\civ} blueshift. In order to investigate the bias in \mbox{\civ-based} \mbh, we measure the \civ\ FWHM from the best-fitting \civ\ profile, as well as the luminosity at 1350 \AA. The \civ-based SMBH mass is calculated following the formula of \cite{VP06}.

\section{Result and Discussion}
\subsection{Blueshift of the whole \civ\ emission line}

\begin{figure*}
\centering
\includegraphics[width=0.5\textwidth]{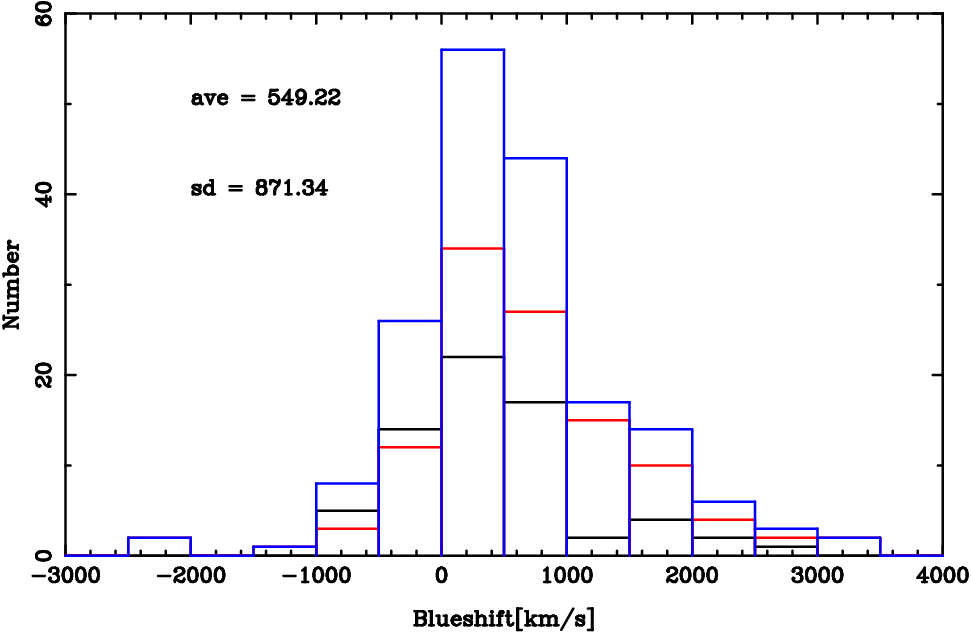}
\caption{The distributions of the velocity shift of the whole \civ\ emission line for PG QSOs (black), high-$z$ QSOs (red) and the entire sample (blue). The average and standard deviation of the distribution are marked on the top right.The positive value denotes the blueshift in units of \kms.}
\label{fig3}
\end{figure*}

Based on the rest frame defined by \oiii $\lambda$5007 or other low-ionization emission line (e.g., \hb), the blueshift of the whole \civ\ profile is investigated for our sample. Figure \ref{fig3} shows the distribution of \civ\ velocity shift for low-$z$ PG QSOs (black), high-$z$ QSOs (red) and our sample (blue). Most objects exhibit blueshift (positive values) and the biggest blueshift is up to $\sim$ 3200 \kms. The average and standard deviation of the distribution are 549.22 \kms and  871.34 \kms respectively. We note that \cite{Coatman2017} showed a more extended tail of blueshift ($>$3000 \kms) based on the rest-frame defined by Balmer lines, which may originate from different definition of systemic redshift in our work and their work. We also note that the blueshift of PG and high-$z$ QSOs are $\sim$ 300 \kms and $\sim$ 700 \kms respectively. The lower blueshift for PG QSOs is correlative with the lower bolometric luminosity, which reduces the average blueshift of our sample.
\begin{figure*}
\centering
\includegraphics[width=0.5\textwidth]{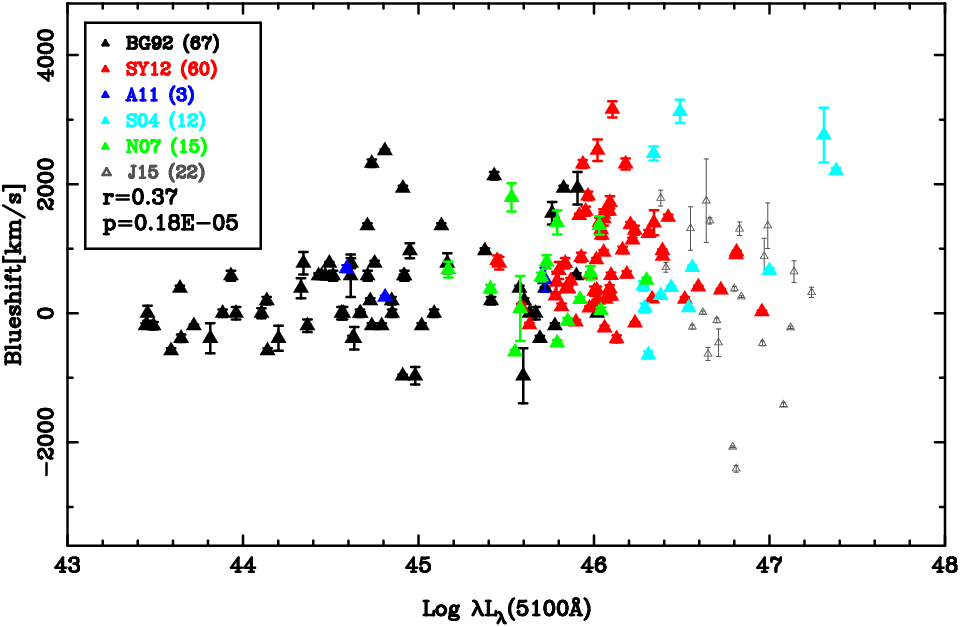}
\caption{The \civ\ velocity shift vs. the luminosity at 5100~\AA\ for our sample. The points with different colors represent that these sources are form different literatures and the number in the brackets indicates the number of objects used here as indicated in the legend}.  A subsample without the sources from \citet[]{Jun2015} (gray points) is used to do the spearman correlation test. The spearman correlation coefficient and the probability of the null hypothesis are 0.37 and $1.8\times 10^{-6}$ respectively.
\label{fig4}
\end{figure*}

\begin{figure*}
\centering
\includegraphics[width=0.5\textwidth]{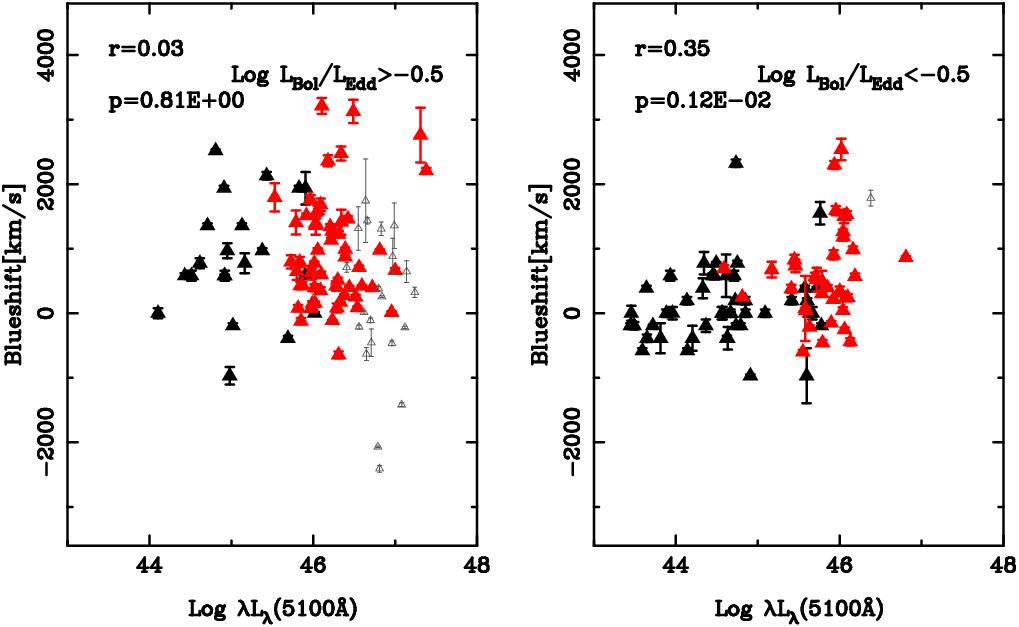}
\caption{The relationship between the \civ\ velocity shift and $L (5100\AA)$ in high \leddR (left) and low \leddR (right). The black and red points represent low-$z$ and high-$z$ QSOs (including \cite{Shen2012}, \cite{Assef2011}, \cite{Shemmer2004} and \cite{Netzer07}) respectively. The gray points are from \citet[]{Jun2015}. The method of calculating spearman correlation coefficients is the same as Figure \ref{fig4}.}
\label{fig5}
\end{figure*}

\begin{figure*}
\centering
\includegraphics[width=0.5\textwidth]{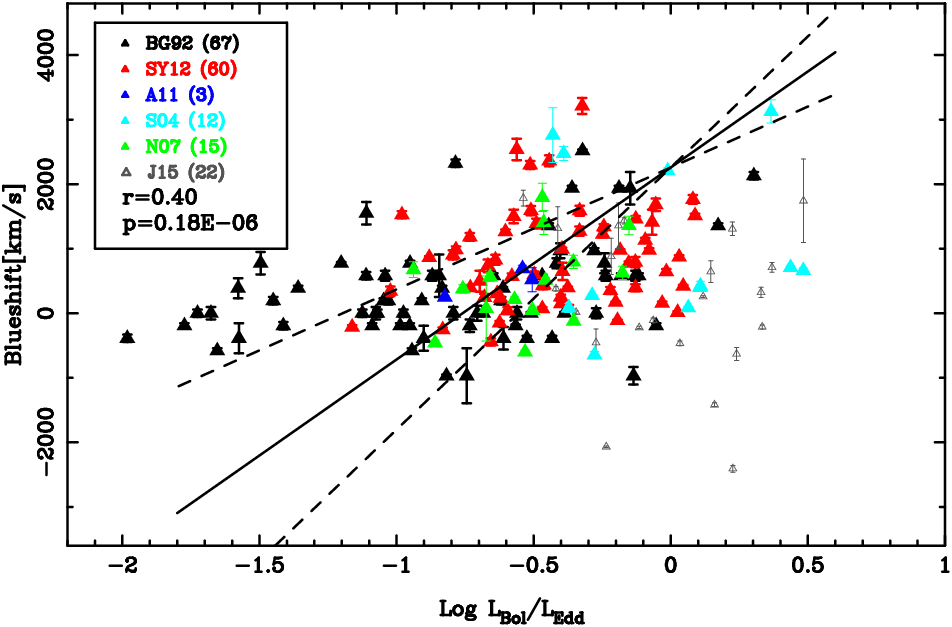}
\caption{The relationship between the \civ\ velocity shift and \leddR. All symbols and labels are the same as Figure \ref{fig4}. The solid line is the best fitting of the BCES bisector (removing the sources from \citet[]{Jun2015}) and the dashed lines are the best fitting of the BCES bisector with $2\sigma$ variance. 0.3 dex is adopted as the typical error of \leddR in the fitting. }
\label{fig6}
\end{figure*}

\begin{figure*}
\centering
\includegraphics[height=6cm]{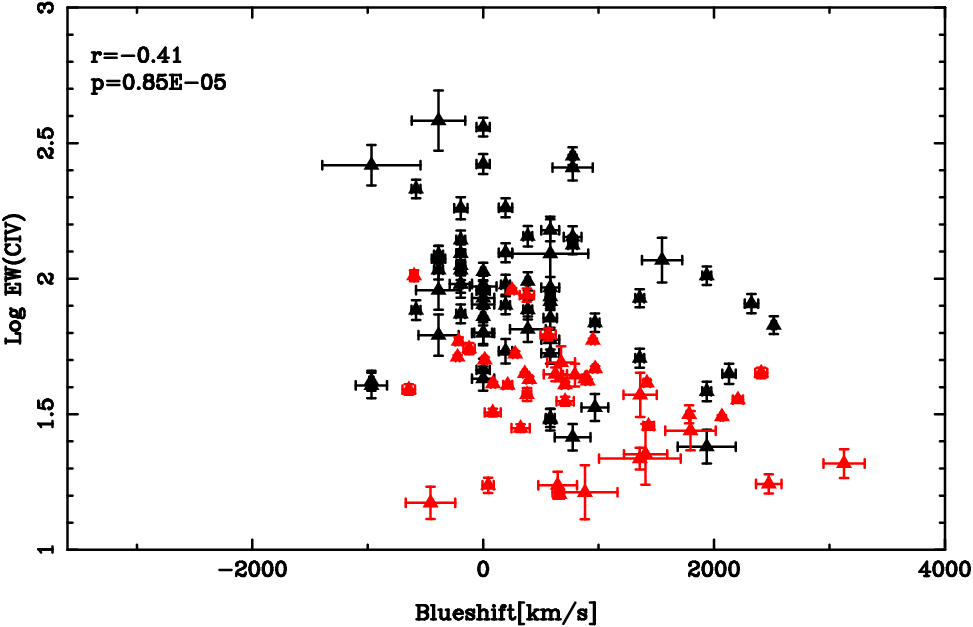}
\caption{The relationship between the \civ\ velocity shfit and EW(\civ). The correlation coefficients are marked on the top of the panel.
PG QSOs and high-$z$ QSOs with \civ\ S/N $>5$ and $z>2.0$ are marked on black and red points.}
\label{fig7}
\end{figure*}

We explore the relationship between the \civ\ velocity shift and the continuum luminosity at 5100\AA. Figure \ref{fig4} shows that there is a weak correlation between the \civ\ blueshift and $L (5100\AA)$ for the sample, where $r=0.23$ and $p=2.3\times 10^{-3}$. The weak correlation may be due to the mixing of \leddR. Figure \ref{fig5} shows the relationship between the \civ\ blueshift and the continuum luminosity at 5100\AA\ in different \leddR bins. We can see the correlation coefficient for low \leddR is higher that of high \leddR (0.38 vs 0.16). The result is consistent with that given by \citet[]{Shen2016}. They investigated the velocity shifts of QSOs emission lines from SDSS reverberation mapping project and found the velocity shift of \civ\ relative to \mgii\ has stronger luminosity dependence than other emission lines (such as \oii, \oiii and \mgii). It is worth noting that our correlation test in right panel of Figure \ref{fig5} agrees with \citet[]{Shen2016} ($r \sim$ 0.4) when considering only the objects with the continuum luminosity less than $\sim 10^{46}$ \ergs.
In this work, there are 14 and 11 radio loudness QSOs in PG QSOs and high-$z$ QSOs respectively. We find the correlation coefficients for the \civ\ blueshift and $L (5100\AA)$/\leddR only increase by 0.06 after excluding the radio loudness QSOs. Therefore,  radio loudness QSOs in our sample do not affect significantly our results. These radio loudness QSOs have, on average, smaller \civ\ blueshift (132 \kms) than that of radio quiet QSOs (616 \kms), which is consistent with the result of \cite{Richards2011}.

Figure \ref{fig6} displays the relationship between the \civ\ blueshift and \leddR for the sample, where $r=0.27$ and $p=3.3\times 10^{-4}$. It is found that the relation will become flat when we go to higher \leddR. 
The possible explanation for the flat relation is that the size of BLR does not depend on the luminosity in high accretion rate QSOs \citep{Du2018}.
Some studies have suggested that the accretion disc will become geometrically thick (e.g., slim disc) in high accretion rate regime, which significantly reduces the amount of the ionizing photons reaching BLR, and then leads to saturated luminosity \citep[]{Wang1999,Wang2014}. The self-shadowing effect of slim discs reduces the radiation pressure and flatten the relation between the \civ\ blueshift and \leddR or $L (5100\AA)$.
\citet[]{Ge2016} estimated the intrinsic total luminosity of \mbox{low-$z$} and \mbox{high-$z$} QSOs based on a simple scaling factor. Such a rough calculation may not reveal the intrinsic relations of these parameters. Actually, the reradiation from dust has great contamination to the intrinsic AGN luminosity, thus the \leddR. \citet[]{Marconi2004} constructed a SED template spectrum without the infrared bump based on $\alpha$ (optical-UV spectral index) and \aox~ (optical-X-ray spectral index). They derived the bolometric correction which depends on monochromatic luminosity from the templates. For our sample, the range of luminosity at 5100 \AA\ is from $10^{43}$ to $10^{47}$ \ergs and the corresponding bolometric correction is between 5 and 9. So in this work, we may overestimate the \leddR, thus affecting our results.
We also find a increasing correlation between the \civ\ blueshift and the luminosity of 5100 \AA\  (\leddR) when we remove the sources from \cite{Jun2015} in our sample. The correlation coefficient and the probability of the null hypothesis are 0.37 (0.40) and $1.8 \times 10^{-6}$ ($1.8\times 10^{-7}$) respectively. 
Simultaneously, we fitted the relationship between the \civ\ blueshift and \leddR (solid line) in Figure \ref{fig6}. The formula for the best fitting is
\begin{eqnarray}
\rm V (\rm CIV)=2257.90(\pm312.17)+2974.10(\pm548.23)\nonumber\\ \rm logL_{\rm bol}/L_{\rm Edd}
\label{eq2}
\end{eqnarray}
The resolution of infrared spectra from \cite{Jun2015} is the lowest (R $\sim$ 200), which can also affect the precision of \leddR. In addition, \cite{Wang2011} presented a comparison of kinematics between \civ\ and \mgii\ emission lines using SDSS data. They found that the blueshift of \civ\ is strongly correlative with \leddR, which is especially prominent in high \leddR regime. The same result was found by \citet[]{Sun2018} recently, who used multi-epoch SDSS spectra to investigate the dependence of the \civ\ blueshift on QSO properties.

Some scenarios have been proposed to explain the physical origin of the \civ\ blueshift. One of them is the orientation effect  
 \cite[]{Denney2012}. However, the scenario was excluded by \citet[]{Runnoe2014}, Who found no correlation between the \civ\ blueshift and the orientation based on a sample of radio core dominance. In the other hand, the large \leddR can cause high-blueshift emission lines, as mentioned above.
In general, the QSOs with high \leddR display more apparent blueshift before approaching saturated luminosity. It indicates that the radiation pressure play an important role in driving the shift of peak wavelength of emission lines.

With the principal component analysis (PCA) of the low-$z$ PG sample, \cite{BG92} found that Principal Component 1 (PC1) is related to the relative strength of optical \feii to \hb ($R_{Fe}$, the ratio between the strength of \feii emission and \hb), Principal Component 2 (PC2) links optical luminosity and \aox. With 
$\hb$-based \mbh, \cite{Boroson2002} suggested that PC1 is mainly correlated with \leddR and PC2 has a strong connection with \mbh and \leddR. We explore the relationship between the \civ\ blueshift and PC1/PC2 and $R_{Fe}$  for PG QSOs. It is found that the \civ\ blueshift has a very weak correlation with PC1 or $R_{Fe}$ but has a medium strength correlation with PC2. The correlation coefficients are -0.11, 0.15 and -0.33 for PC1, $R_{Fe}$ and PC2, respectively. The relation between PC1/PC2 and \leddR/\mbh needs to be investigated with more reliable measurements of \mbh in QSOs.

In order to explore the relationship between EW (\civ) and the \civ\ blueshift, We choose a subsample containing 67 PG QSOs and 43 high-$z$ QSOs from our sample. These high-$z$ QSOs have \civ\ emission line S/N ratio $>5$ and $z>2$. Figure \ref{fig7} shows the relationship between EW (\civ) and the \civ\ blueshift. A medium strength correlation is found between them ($r=-0.41$), which is consistent with the wind-disk model of BLRs given by \cite{Richards2011}. Higher \leddR may lead to the formation of a wind component. If the BLR is dominated by the wind component, broad emission lines will show the blueshift and their strength will be suppressed \citep[]{Richards2011}. 

Alternatively, \cite{Shemmer2015} pointed out that the lowest scatter relationship is not continuum vs. EW (\civ), but rather $\hb$-based \leddR, i.e., modified Baldwin effect (MBE). The MBE can also be used to explain the relationship between the total EW (\civ) and the \civ\ blueshift. As \leddR changes, there are expected changes in the thickness of the inner accretion disk, which may well affect the ionization of the BLR and a disk wind. Furthermore, if objects are viewed from the direction of pole-on, then these objects will exhibit narrower line width, but actually lower \leddR (lower \civ\ blueshift). Therefore orientation effect cannot be the main explanation of the outliers in Figure \ref{fig7}. Weak line QSOs deviating considerably from the classical Baldwin effect and MBE in \cite{Shemmer2015} reveal that the relationship between the profile and EW (\civ) may not only depend on \leddR, but also additional physical properties (such as BLR geometry, density and metallicity).

The relationship between EW (\civ) and the \civ\ blueshift also supports the classical Baldwin effect that originating from the softening of high-energy photons \citep{Netzer1992,Korista1998,Dietrich2002}. The larger blueshift indicates stronger wind component. As suggested by \cite{Richards2011}, the reduction of the number of high-energy photons that can ionize gas is beneficial to the formation of wind component, thus enhancing the blueshift. The wind component is not separated from the disk component in our fitting model. Therefore we are not sure that whether the Baldwin effect is triggered by the disappearance or weakness of wind component.

\subsection{\civ-based \mbh correction with the \civ\ blueshift}

\begin{figure*}
\centering
\includegraphics[height=6cm]{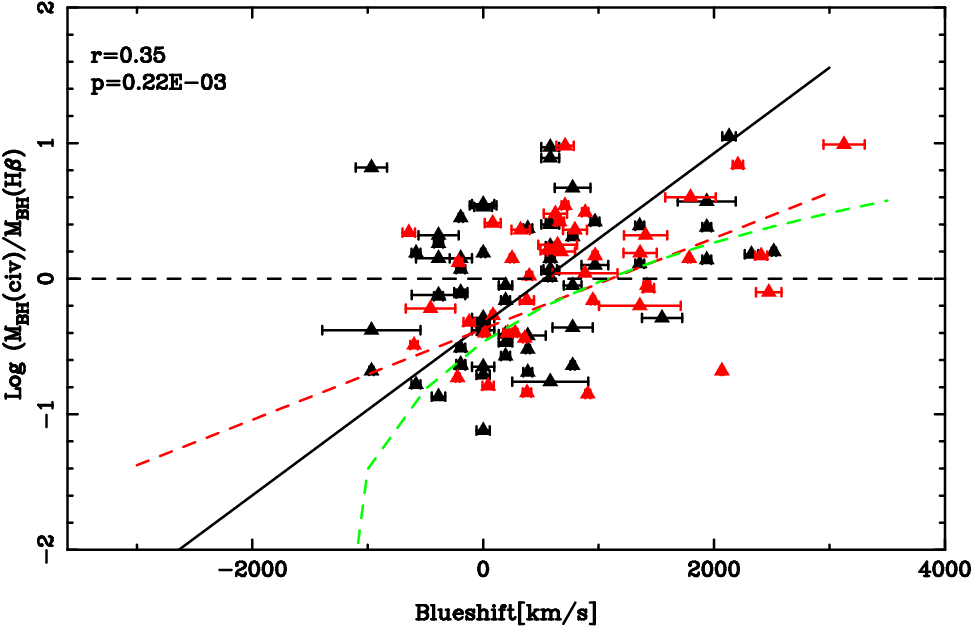}
\caption{The effect of \civ\ velocity shift on the difference between \mbh based on \civ\ and \hb. The black solid line is the best fitting of the BCES bisector. Red and green dashed lines are the relations from \citet[]{Jun2017} and \citet[]{Coatman2017}. The correlation coefficients are listed on the top of the panel. Objects used for the analysis are the same as in Figure \ref{fig7}.}
\label{fig8}
\end{figure*}

It is found that the \civ-based \mbh is biased with respect to that based on \hb \cite[]{Shen2012, Bian2012}. This bias is suggested to be corrected by the \civ\ blueshift \citep[]{Jun2017, Coatman2017}. Figure \ref{fig8} shows the relationship between the \mbh difference (\civ-based \mbh and $\hb$-based \mbh) and the \civ\ blueshift. It is found that there is a medium strong correlation between them with the correlation coefficient of $r=0.35$ and the probability of the null hypothesis of $p=2.2\times 10^{-4}$. The black line is the best fit of the BCES bisector. The formula is as follows:
\begin{eqnarray}
\rm log (M_{\rm BH}(\rm {CIV/H\beta}))=-0.34(\pm0.09)+6.31(\pm1.57)\nonumber\\
\times 10^{-4} \rm V(CIV)
\label{eq2}
\end{eqnarray}
The red and green dashed lines in Figure \ref{fig8} are from \cite{Jun2017} and \cite{Coatman2017}, respectively. We correct the \civ-based \mbh according our best fit and find that the mean value and the dispersion of the \mbh difference are 0.012 and 0.48 dex respectively, which shows that some other parameters need to be considered in the calculation of \mbh from the \mbox{\civ} line.

Compared to H$\beta$-based \mbh, many previous studies have shown that the \civ-based single-epoch SMBH usually exhibits significant scatter \citep{Baskin2004,Richards2011,Denney2012,Bian2012,Shen2012}. Recent study shown by \cite{Coatman2017} used a high-$z$ AGN sample ($1.5<z<4.0$) to investigate the relationship between the \civ\ blueshift and \civ-based \mbh. They found that \civ-based \mbh may be a factor 5 to 10 larger than that of Balmer-line-based at the \civ\ blueshift of more than 3000 \kms and given an empirical correction for the \civ-based \mbh based on the relation between the \civ\ blueshift and the ratio of \civ\ FWHM to H$\alpha$ FWHM. In contrast, \cite{Mejia2018} used a larger sample (including the objects from \cite{Coatman2017}) and shown that the relation between FWHM (\civ/H$\alpha$)/FWHM (\civ/\mgii) and the \civ\ blueshift is driven by the relation between the \civ\ blueshift and \civ\ FWHM, which suggested that the empirical correction constructed by the \civ\ blueshift and Balmer-line width for \civ-based single-epoch \mbh, therefore, the \leddR,  is limited. In addition, they also found that there is no connection between \mgii\ and \civ\ profiles according to the PCA analysis, suggesting high-ionization (\civ) and low-ionization (\ha, \mgii) lines are distinct.

We find the \civ\ blueshift is indeed related to the ratio of \mbh (Figure \ref{fig8}). However, our correction for \civ-based single-epoch \mbh is limited considering the large system error of \mbh, which is also seems to support the conclusion given by \cite{Mejia2018} that using solely \civ\ line width  may not get the reliable virialized \mbh. Line peak ratios needs to be considered \citep{Runnoe2013,Brotherton2015}. To understand how to improve the accuracy of  \civ-based single-epoch \mbh, large sample including low- and high-$z$ AGNs is required to analyze in the future.

The difference of the results among \cite{Coatman2017}, \cite{Mejia2018} and us may originate from the sample selection. 
The similarity about the samples is that all these studies use the sample of \cite{Shen2012}. However, most of the objects in \cite{Shen2012} are $z<1.7$, which leads the incompleteness of the \civ\ profiles in blue waveband. Therefore, the profile parameters, such as FWHM and EW may be inaccurate for these objects. We select objects with $z>2.0$ to avoid the issue and investigate the effects of the \civ\ blueshift on \civ-based single-epoch \mbh in Section 3.2. In addition, a difference in the samples is that the utilization of PG QSOs ($z<0.5$) in this work. PG QSOs have lower bolometric luminosity than high-$z$ QSOs ($z>2.0$), which allows us to expand our results to the end of low luminosity.

\section{CONCLUSION}
For a collected sample of low-$z$ and high-$z$ QSOs, spectral decomposition is used to investigate the blueshift of \civ\ broad emission line relative to systemic which is defined by \oiii $\lambda$5007. The results are as follows:

(1) It is confirmed that a blueshift exists for the high-ionization \civ\ broad emission line in the rest frame defined by the narrow \oiii line or \hb. It is found that there is a medium strong positive correlation between the \civ\ blueshift and the luminosity of continuum or \leddR, and a medium negative relationship between the \civ\ blueshift and EW(\civ). These results are consistent with the picture that  radiation pressure is correlative to the \civ\ blueshift.

(2) There is a medium strong correlation between the ratio of \civ-based \mbh to $\hb$-based \mbh and the \civ\ blueshift. This relationship depends on the accurate systemic redshift and the $\hb$-based \mbh. A larger sample is needed to investigate this relationship.

\acknowledgments

We would like to thank Michael S Brotherton for \mbox{useful} discussions. This work is supported by the National Science Foundations of China (Nos. 11373024, 11233003 and 11873032).

\begin{deluxetable*}{llllllllllllllllllllllllllll}
\setcounter{table}{1}
\tablenum{1}
\tablecaption{A summary of PG QSOs.}\label{Table 1}
\tablewidth{0pt}
\tabletypesize{\scriptsize}
\tablehead{
\colhead{\#} & \colhead{Name} & 
\colhead{$z$} & \colhead{Velocity shift} & 
\colhead{Error of velocity shift} &
\colhead{$L_{5100}^{\dag}$} & 
\colhead{$\mbh$$^{\dag}$} &
\colhead{$L_{\rm bol}/L_{\rm Edd}^{\dag}$} \\ 
\colhead{} & \colhead{} & \colhead{} & \colhead{(\kms)} &  \colhead{\kms} & \colhead{(\ergs)}  & \colhead{$\msun$} &
\colhead{} \\
(1) & (2)& (3)& (4)& (5)& (6) & (7)& (8)\\
}

\startdata
1     &    $0003+158^{*}$    &   0.450    &        0  &   19  &   46.02  &   9.273 &   -0.388     \\
2     &    $0003+199^{*}$    &   0.025    &     -194  &   19  &   43.72  &   7.152 &   -0.566     \\
3     &    0007+106          &   0.089    &     -194  &   39  &   44.73  &   8.689 &   -1.089     \\
4     &    $0026+129^{*}$    &   0.142    &     -968  &   19  &   44.91  &   8.594 &   -0.818     \\
5     &    0049+171          &   0.064    &     -387  &  232  &   43.81  &   8.254 &   -1.575     \\
6     &    $0050+124^{*}$    &   0.061    &     1356  &   39  &   44.71  &   7.402 &    0.173     \\
7     &    $0052+251^{*}$    &   0.155    &      775  &   39  &   44.75  &   8.567 &   -0.951     \\
8     &    0157+001          &   0.164    &     1937  &   39  &   44.91  &   8.137 &   -0.360     \\
\enddata
\end{deluxetable*}

\setcounter{table}{0}
\begin{deluxetable*}{lllllllllll}
\setcounter{table}{1}
\tablenum{1}
\tablecaption{-- continue}\label{Table 1}
\tablewidth{0pt}
\tabletypesize{\scriptsize}
\tablehead{
\colhead{\#} & \colhead{Name} & 
\colhead{$z$} & \colhead{Velocity shift} & 
\colhead{Error of velocity shift} &
\colhead{$L_{5100}^{\dag}$} & 
\colhead{$\mbh$$^{\dag}$} &
\colhead{$L_{\rm bol}/L_{\rm Edd}^{\dag}$}\\ 
\colhead{} & \colhead{} & \colhead{} & \colhead{(\kms)} &  \colhead{\kms} & \colhead{(\ergs)}  & \colhead{$\msun$} &
\colhead{} \\
(1) & (2)& (3)& (4)& (5)& (6) & (7)& (8)\\
}

\startdata
51    &    $1444+407^{*}$    &   0.267    &      775  &  155  &   45.16  &   8.273 &   -0.242     \\
52    &    1501+106          &   0.036    &      194  &   58  &   44.13  &   8.451 &   -1.450     \\
53    &    $1512+370^{*}$    &   0.371    &      194  &   39  &   45.60  &   9.376 &   -0.908     \\
54    &    1519+226          &   0.137    &      775  &   77  &   44.62  &   7.897 &   -0.416     \\
55    &    1534+580          &   0.030    &     -194  &   39  &   43.45  &   8.085 &   -1.774     \\
56    &    $1543+489^{*}$    &   0.400    &     2130  &   58  &   45.43  &   7.995 &    0.303     \\
57    &    $1545+210^{*}$    &   0.266    &      194  &   58  &   45.41  &   9.309 &   -1.030     \\
58    &    $1612+261^{*}$    &   0.131    &     -387  &   39  &   44.62  &   8.014 &   -0.525     \\
\enddata
\tablecomments{Col.(1): Sequence number;
Col.(2): Name of PG QSOs, $\ast$ in the Col. (2) indicate the observation are from HST;
Col.(3): Redshift from \citet[]{Schmidt1983};
Col.(4-5): Shift of \civ\ peak wavelength and the error, positive value means blueshift;
Col.(6): The luminosity of 5100~\AA;
Col.(7-8): The host-corrected $\mbh$ and Eddington ratio (in logarithm scale).
$\dag$ means that the data is from \citet[]{Ge2016}.}
\end{deluxetable*}

\begin{deluxetable*}{llllllllllllllllllllllllll}
\tablenum{2}
\tablecaption{A summary of high-$z$ QSOs.}\label{Table 2}
\tablewidth{100pt}
\tabletypesize{\scriptsize}
\tablehead{
\colhead{\#} & \colhead{Name} & 
\colhead{$z_{\rm fit}$}& \colhead{Velocity shift} & 
\colhead{Error of velocity shift} &
\colhead{$L_{5100}^{\dag}$} & 
\colhead{$\mbh$$^{\dag}$} &
\colhead{$L_{\rm bol}/L_{\rm Edd}^{\dag}$} &
\colhead{Ref} \\ 
\colhead{} & \colhead{} & \colhead{}& \colhead{(\kms)} & \colhead{(\kms)} & \colhead{(\ergs)} & \colhead{$\msun$} &
\colhead{} & \colhead{} \\
(1) & (2)& (3)& (4)& (5)& (6)& (7) & (8)& (9)\\
}

\startdata
1    &   SDSS225800.02-084143  &  1.496  &     793  &     80  &  45.836   &     8.835   &    -0.129  &  a  \\
2    &   SDSS035856.73-054023  &  1.506  &     649  &    133  &  45.800   &     9.065   &    -0.395  &  a  \\
3    &   SDSS133321.90+005824  &  1.511  &    -152  &     30  &  45.897   &     9.391   &    -0.624  &  a  \\
4    &   SDSS081331.28+254503  &  1.513  &      14  &     19  &  46.957   &     9.802   &     0.025  &  a  \\
5    &   SDSS152111.86+470539  &  1.517  &      71  &     25  &  45.973   &     9.312   &    -0.469  &  a  \\
6    &           FBQ1633+3134  &  1.522  &     519  &    185  &  45.720   &     9.096   &    -0.506  &  b  \\
7    &   SDSS085543.26+002908  &  1.525  &     303  &     30  &  45.781   &     9.284   &    -0.633  &  a  \\
8    &   SDSS123355.21+031327  &  1.528  &     904  &     71  &  45.927   &     9.591   &    -0.794  &  a  \\
\enddata
\end{deluxetable*}

\setcounter{table}{0}
\begin{deluxetable*}{lllllllllllllllllllll}
\tablenum{2}
\tablecaption{-- continue}\label{Table 2}
\tablewidth{100pt}
\tabletypesize{\scriptsize}
\tablehead{
\colhead{\#} & \colhead{Name} & 
\colhead{$z_{\rm fit}$}& \colhead{Velocity shift} & 
\colhead{Error of velocity shift} &
\colhead{$L_{5100}^{\dag}$} & 
\colhead{$\mbh$$^{\dag}$} &
\colhead{$L_{\rm bol}/L_{\rm Edd}^{\dag}$} &
\colhead{Ref} \\ 
\colhead{} & \colhead{} & \colhead{}& \colhead{(\kms)} & \colhead{(\kms)} & \colhead{(\ergs)} & \colhead{$\msun$} &
\colhead{} & \colhead{} \\
(1) & (2)& (3)& (4)& (5)& (6)& (7) & (8)& (9)\\
}

\startdata
61   &   SDSS143148.09+053558  &  2.095  &     863  &     22  &  46.810   &    10.561   &    -0.881  &  a  \\
62   &   SDSS142108.71+224117  &  2.188  &     976  &     17  &  46.813   &     9.868   &    -0.185  &  a  \\
63   &                 TON618  &  2.219  &    2761  &    423  &  47.310   &    10.610   &    -0.430  &  c  \\
64   &                  UM645  &  2.268  &    -644  &     54  &  46.310   &     9.458   &    -0.278  &  c  \\
65   &   SDSSJ170102.18+61230  &  2.290  &    2474  &    110  &  46.340   &     9.601   &    -0.391  &  c  \\
66   &   SDSSJ115111.20+03404  &  2.337  &      74  &    504  &  45.580   &     9.123   &    -0.673  &  d  \\
67   &   SDSSJ144245.66-02425  &  2.355  &    1360  &    144  &  46.030   &     9.052   &    -0.152  &  d  \\
68   &   SDSSJ100710.70+04211  &  2.367  &     675  &    121  &  45.170   &     8.978   &    -0.938  &  d  \\
\enddata
\tablecomments{Col.(1): Sequence number;
Col.(2): Name of high-$z$ QSOs;
Col.(3): Redshift used to fit spectrum;
Col.(4-5): Shift of \civ\ peak wavelength and the error, positive value means blueshift;
Col.(6): The luminosity of 5100~\AA;
Col.(7-8): $\mbh$ and Eddington ratio (in logarithm scale);
Col.(9): The references. a: \citet[]{Shen2012}, b; \citet[]{Assef2011}, c: \citet[]{Shemmer2004}, d: \citet[]{Netzer07}, e: \citet[]{Jun2015}.
$\dag$ means that the data is from \citet[]{Ge2016} and the fitting redshifts of UM667 and TON618 are from NASA/IPAC Extragalactic Database.}
\end{deluxetable*}



\begin{deluxetable*}{llllllllllllllllllllllllllllll}
\setcounter{table}{2}
\tablecaption{A summary of fitting result for PG QSOs.}\label{Table 3}
\tablewidth{0pt}
\tabletypesize{\scriptsize}
\tablehead{
\colhead{\#} & \colhead{Name} & 
\colhead{$z_{fit}$} & \colhead{$\rm L_{N}$} & 
\colhead{$\rm FWHM_{N}$} &
\colhead{$\rm peak_{N}$} & 
\colhead{$\rm  L_{B}$} &
\colhead{$\rm  FWHM_{b}$} &
\colhead{$\rm  peak_{B}$} \\ 
\colhead{} & \colhead{} & \colhead{} & \colhead{(\ergs)} &  \colhead{(\kms)} & \colhead{(\kms)}  &  \colhead{(\ergs)} &  \colhead{(\kms)} & \colhead{(\kms)} &
\colhead{} \\
(1) & (2)& (3)& (4)& (5)& (6) & (7)& (8) & (9)\\
}

\startdata
1    &   0003+158   &  0.450   &  44.28$\pm$0.01   &  2558$\pm$  58   &     5$\pm$ 20   &  44.67$\pm$0.01   & 12357$\pm$223   &  -847$\pm$ 72  \\ 
2    &   0003+199   &  0.025   &  42.70$\pm$0.01   &  2349$\pm$  37   &  -223$\pm$ 14   &  42.58$\pm$0.01   & 14492$\pm$619   &   192$\pm$188  \\ 
3    &   0007+106   &  0.089   &  42.97$\pm$0.02   &  3003$\pm$ 122   &  -170$\pm$ 41   &  43.17$\pm$0.02   & 15174$\pm$026   &  -509$\pm$277  \\ 
4    &   0026+129   &  0.142   &  42.93$\pm$0.02   &  1536$\pm$  69   & -1006$\pm$ 27   &  43.50$\pm$0.01   & 14231$\pm$590   &  -932$\pm$187  \\ 
5    &   0049+171   &  0.064   &  42.81$\pm$0.14   &  4556$\pm$ 816   &  -386$\pm$231   &  42.72$\pm$0.16   & 10520$\pm$282   &   113$\pm$501  \\ 
6    &   0050+124   &  0.061   &  42.61$\pm$0.02   &  2457$\pm$  87   &  1369$\pm$ 34   &  42.80$\pm$0.02   & 20000$\pm$521   &  2905$\pm$442  \\ 
7    &   0052+251   &  0.155   &  43.66$\pm$0.02   &  4134$\pm$ 147   &   823$\pm$ 43   &  43.98$\pm$0.01   & 14621$\pm$475   &   207$\pm$120  \\ 
8    &   0157+001   &  0.164   &  43.38$\pm$0.02   &  3688$\pm$ 143   &  1872$\pm$ 41   &  43.57$\pm$0.02   & 20000$\pm$488   &  2675$\pm$337  \\
\enddata
\end{deluxetable*}

\begin{deluxetable*}{lllllllllllllllll}
\setcounter{table}{2}
\tablecaption{-- continue}\label{Table 3}
\tablewidth{0pt}
\tabletypesize{\scriptsize}
\tablehead{
\colhead{\#} & \colhead{Name} & 
\colhead{$z_{fit}$} & \colhead{$\rm L_{N}$} & 
\colhead{$\rm FWHM_{N}$} &
\colhead{$\rm peak_{N}$} & 
\colhead{$\rm L_{B}$} &
\colhead{$\rm FWHM_{b}$} &
\colhead{$\rm peak_{B}$} \\ 
\colhead{} & \colhead{} & \colhead{} & \colhead{(\ergs)} &  \colhead{(\kms)} & \colhead{(\kms)}  &  \colhead{(\ergs)} &  \colhead{(\kms)} & \colhead{(\kms)} &
\colhead{} \\
(1) & (2)& (3)& (4)& (5)& (6) & (7)& (8) & (9)\\
}

\startdata
51   &   1444+407   &  0.267   &  43.04$\pm$0.15   &  2179$\pm$ 507   &   748$\pm$159   &  43.81$\pm$0.03   &  7761$\pm$508   &  1440$\pm$173  \\ 
52   &   1501+106   &  0.036   &  42.50$\pm$0.03   &  3389$\pm$ 160   &   141$\pm$ 52   &  42.80$\pm$0.01   & 13579$\pm$586   &   617$\pm$162  \\ 
53   &   1512+370   &  0.371   &  43.79$\pm$0.02   &  2595$\pm$ 117   &    68$\pm$ 41   &  44.38$\pm$0.01   & 12778$\pm$287   &  1001$\pm$ 96  \\ 
54   &   1519+226   &  0.137   &  42.96$\pm$0.04   &  2432$\pm$ 188   &   754$\pm$ 68   &  43.23$\pm$0.03   & 14662$\pm$466   &  -218$\pm$457  \\ 
55   &   1534+580   &  0.030   &  42.02$\pm$0.03   &  1844$\pm$ 114   &  -289$\pm$ 40   &  42.59$\pm$0.01   &  8679$\pm$247   &   323$\pm$ 87  \\ 
56   &   1543+489   &  0.400   &  43.80$\pm$0.03   &  4802$\pm$ 200   &  1997$\pm$ 55   &  43.81$\pm$0.03   & 15044$\pm$147   &  2905$\pm$281  \\ 
57   &   1545+210   &  0.266   &  43.84$\pm$0.03   &  3830$\pm$ 177   &   230$\pm$ 58   &  44.21$\pm$0.01   & 15845$\pm$617   &   302$\pm$160  \\ 
58   &   1612+261   &  0.131   &  43.06$\pm$0.03   &  2781$\pm$ 137   &  -285$\pm$ 47   &  43.38$\pm$0.01   & 10543$\pm$379   & -1724$\pm$153  \\
\enddata
\tablecomments{Col.(1): Sequence number;
Col.(2): Name of PG QSOs;
Col.(3): Redshift from \citet[]{Schmidt1983};
Col.(4-9): The profile fitting parameters of \civ\ (luminosity, FWHM and peak). Subscript N and B indicate narrow and broad components respectively.}
\end{deluxetable*}

\begin{deluxetable*}{lllllllllllllllllllllllll}
\tablecaption{A summary of fitting result for high-$z$ QSOs.}\label{Table 4}
\tablewidth{100pt}
\tabletypesize{\scriptsize}
\tablehead{
\colhead{\#} & \colhead{Name} & 
\colhead{$z_{fit}$} & \colhead{$\rm L_{N}$} & 
\colhead{$\rm FWHM_{N}$} &
\colhead{$\rm peak_{N}$} & 
\colhead{$\rm L_{B}$} &
\colhead{$\rm FWHM_{b}$} &
\colhead{$\rm peak_{B}$} \\ 
\colhead{} & \colhead{} & \colhead{} & \colhead{(\ergs)} &  \colhead{(\kms)} & \colhead{(\kms)}  &  \colhead{(\ergs)} &  \colhead{(\kms)} & \colhead{(\kms)} &
\colhead{} \\
(1) & (2)& (3)& (4)& (5)& (6) & (7)& (8) & (9)\\
}

\startdata
1     &   SDSS225800.02-084143   &  1.496   &  45.54$\pm$0.05   &  3753$\pm$ 226  &  807$\pm$ 80   &  46.00$\pm$0.07   & 16574$\pm$ 1581   &  2905$\pm$131  \\
2     &   SDSS035856.73-054023   &  1.506   &  45.04$\pm$0.11   &  3454$\pm$ 446  & 1007$\pm$133   &  45.24$\pm$0.08   &  8050$\pm$  676   & -1079$\pm$614  \\
3     &   SDSS133321.90+005824   &  1.511   &  45.64$\pm$0.02   &  2794$\pm$  86  &  361$\pm$ 30   &  46.25$\pm$0.01   & 12624$\pm$  247   &  -910$\pm$ 80  \\
4     &   SDSS081331.28+254503   &  1.513   &  46.42$\pm$0.02   &  3394$\pm$  83  &  843$\pm$ 19   &  46.48$\pm$0.01   & 10223$\pm$  406   &  1349$\pm$ 92  \\
5     &   SDSS152111.86+470539   &  1.517   &  45.73$\pm$0.02   &  2664$\pm$  74  &  409$\pm$ 25   &  45.98$\pm$0.01   & 12977$\pm$  565   &    21$\pm$163  \\
6     &           FBQ1633+3134   &  1.522   &  45.32$\pm$0.12   &  2265$\pm$ 256  & 2905$\pm$185   &  45.69$\pm$0.05   &  3073$\pm$  239   &   571$\pm$175  \\
7     &   SDSS085543.26+002908   &  1.525   &  45.43$\pm$0.02   &  2664$\pm$  93  &  149$\pm$ 30   &  45.82$\pm$0.01   & 12712$\pm$  494   &   770$\pm$143  \\
8     &   SDSS123355.21+031327   &  1.528   &  45.41$\pm$0.05   &  4380$\pm$ 261  &  954$\pm$ 71   &  45.99$\pm$0.01   & 14673$\pm$  693   &  1088$\pm$164  \\
\enddata
\end{deluxetable*}

\setcounter{table}{3}
\begin{deluxetable*}{llllllllllllllllllllllllll}
\tablecaption{-- continue}\label{Table 4}
\tablewidth{100pt}
\tabletypesize{\scriptsize}
\tablehead{
\colhead{\#} & \colhead{Name} & 
\colhead{$z_{fit}$} & \colhead{$\rm L_{N}$} & 
\colhead{$\rm FWHM_{N}$} &
\colhead{$\rm peak_{N}$} & 
\colhead{$\rm L_{B}$} &
\colhead{$\rm FWHM_{b}$} &
\colhead{$\rm peak_{B}$} \\ 
\colhead{} & \colhead{} & \colhead{} & \colhead{(\ergs)} &  \colhead{(\kms)} & \colhead{(\kms)}  &  \colhead{(\ergs)} &  \colhead{(\kms)} & \colhead{(\kms)} &
\colhead{} \\
(1) & (2)& (3)& (4)& (5)& (6) & (7)& (8) & (9)\\
}

\startdata
61    &   SDSS143148.09+053558   &  2.095   &  46.54$\pm$0.01   &  3676$\pm$  73  &  445$\pm$ 22   &  46.82$\pm$0.01   & 14190$\pm$  293   &   348$\pm$ 77  \\
62    &   SDSS142108.71+224117   &  2.188   &  46.88$\pm$0.01   &  4322$\pm$  51  &  655$\pm$ 17   &  46.83$\pm$0.01   & 14387$\pm$  329   &   944$\pm$ 78  \\
63    &                 TON618   &  2.219   &  45.44$\pm$0.35   &  3083$\pm$1398  & 1331$\pm$423   &  46.66$\pm$0.02   &  8620$\pm$  362   &  2905$\pm$179  \\
64    &                  UM645   &  2.268   &  46.35$\pm$0.03   &  4677$\pm$ 202  &  360$\pm$ 54   &  46.35$\pm$0.02   & 18706$\pm$ 2100   &   172$\pm$480  \\
65    &   SDSSJ170102.18+61230   &  2.290   &  45.99$\pm$0.05   &  4196$\pm$ 286  & 1437$\pm$110   &  46.03$\pm$0.05   & 13274$\pm$ 1328   &  2905$\pm$466  \\
66    &   SDSSJ115111.20+03404   &  2.337   &  45.21$\pm$0.68   &  1847$\pm$ 602  &  292$\pm$504   &  45.18$\pm$0.75   &  2483$\pm$ 1817   & -1139$\pm$825  \\
67    &   SDSSJ144245.66-02425   &  2.355   &  45.80$\pm$0.11   &  4455$\pm$ 585  & 1322$\pm$144   &  45.79$\pm$0.12   & 10632$\pm$ 1897   &   369$\pm$595  \\
68    &   SDSSJ100710.70+04211   &  2.367   &  45.79$\pm$0.07   &  4727$\pm$ 448  & 1054$\pm$121   &  45.64$\pm$0.09   & 14292$\pm$ 3832   & -1102$\pm$471  \\
\enddata
\tablecomments{Col.(1): Sequence number;
Col.(2): Name of high-$z$ QSOs;
Col.(3): Redshift used to fit spectrum;
Col.(4-9): The profile fitting parameters of \civ\ (luminosity, FWHM and peak). Subscript N and B indicate narrow and broad components respectively.
Note that the profile parameters may be problematic for QSQs with $z<1.7$ due to the incompleteness of \civ\ profile.}
\end{deluxetable*}

\end{document}